# Data Warehouse Success and Strategic Oriented Business Intelligence: A Theoretical Framework

Eiad Basher Alhyasat

Faculty of Planning and Management, Al-Balqa Applied University

Al-Salt, Jordan

E-mail: Eiadb65@hotmail.com

Mahmoud Al-Dalahmeh

Faculty of Business, The University of Jordan

Amman, Jordan

E-mail: m.aldalahmeh@ju.edu.jo

## Abstract

With the proliferation of the data warehouses as supportive decision making tools, organizations are increasingly looking forward for a complete data warehouse success model that would manage the enormous amounts of growing data. It is therefore important to measure the success of these massive projects. While general IS success models have received great deals of attention, few research has been conducted to assess the success of data warehouses for strategic business intelligence purposes. The framework developed in this study consists of the following nine measures: Vendors and Consultants, Management Actions, System Quality, Information Quality, Data Warehouse Usage, Perceived utility, Individual Decision Making Impact, Organizational Decision Making Impact, and Corporate Strategic Goals Attainment.

## 1. Introduction

In today's highly competitive business climate, initiating high quality decisions can be either the choice of surviving or thriving. The high risk/high return management of Data Warehouses (DW) (Watson et al., 2004; Al-Debei, 2011) is a complex undertaking, since the challenge of the phenomenal growth of high volumes of data can be risky in terms of cost effectiveness and security manners (Ang and Teo, 2000; Chen et al., 2000). Therefore, Information Technologies such as Data warehouses can fall short and be a source of losses and concerns to organizations if not strategically planned (Subramanian et al., 1997; Heo and Han, 2003).

Since the 1990s, it has been quite remarkable how fast the Data Warehouse (DW) market has grown (McFadden, 1996; Ramamurthy et al., 2008). Effortlessly, organizations are trying to create a competitive edge through leveraging their source of data in decision making for strategic intelligence purposes (Srivastava and Chen, 1999; Samtani et al., 1999; Chen et al., 2000; Bruckner et al., 2001; Watson et al., 2002; Chau et al., 2002; Hwang et al., 2004; Park, 2006; Al-Debei, 2011). A data warehouse is a queryable source of highly organized cross functional data for the intention of enhancing problem identification and persuading critical management decision needs (Chen et al., 2000; Bruckner et al., 2001; March and Hevner, 2007). Simply put, as McFadden stated: "the DW is a database that is optimized for decision support"; that explains the extensive acquisitions of these centralized data-driven repositories as basis to serve the need for highly integrated enterprise-wide information (Bruckner et al., 2001; March and Hevner, 2007). However, organizations are eager to cut down the negative impacts of these huge IT investments, and therefore keen to measure the illusive factors behind successful information systems (Petter et al., 2008).

Despite that the IS field has strived to understand the nature and meaning of information systems success, as the popularity of DeLone and McLean IS success model proves that (DeLone and McLean, 1992; 2002, 2003; 2004; Seddon and Kiew, 1997; Saarinen, 1996; Roldan and Leal, 2003; Jennex and Olfman, 2003; Wu and Wang, 2006; Petter et al., 2008; Al-Debei et al., 2013; Al-Debei, 2013). Yet scarce literature has contributed to measure the endurance success of data warehouse projects for decision making purposes. Noticeably, there is still the need for a careful and balanced discussion for capturing strategic business value of business intelligence through leveraging data warehouse investments. As a result, the significance of this piece of work seeks to exploit what is beyond the traditional success of data warehouse systems. The approach of developing a more comprehensive framework taking into consideration the 'cause and effect' relationships of the DeLone and McLean IS success model, along with other factors related to the Information Systems Implementation (ISI) model (Land, 1994). Rather than considering the same IS success model of DeLone and McLean (2002) to support the success constructs customized in this paper like most researchers tended to do (DeLone and McLean, 2003). We aim to build up a complete model for strategic business intelligence success, with a motivation for a more in depth analytical research for strategic goals attainment purposes (i.e. leveraging DW for decision making and business intelligence purposes). This study attempts to clarify the confusion of what pre/post constructs best measure data warehouse projects for decision making and business intelligence processes. Until very recently, the entire literature on IS in developing countries has been ignored, and the focus of researchers and practitioners has been elsewhere. Therefore, this study aims to help in explaining why so many information systems in a developing country like Jordan would fail.

## 2. Background and Literature Review

### 2.1 Data Warehouse (DW) and Business Intelligence (BI)

Since the early 1990s, data warehouses have been at the forefront of information technology applications as a way for organizations to effectively use digital information for business planning and decision making. As researchers, we no doubt will encounter the data warehouse phenomenon; hence, an understanding of data warehouse system architecture is or will be important in roles and responsibilities of information management. A DW is considered one of the most powerful decision support and business intelligence technologies that have emerged in the last decade (Ramamurthy et al., 2008; Al-Debei, 2011). Nevertheless, the realization of DW benefits by business organizations has been below expectations (Watson et al., 2002). Hence, this study focuses on two aspects: first, it discusses the role and value of DW as an aspect or driver for business intelligence, and secondly it is critically analyzing both organizational and technological issues affecting the process of leveraging data warehouse success for strategic oriented purposes.

Data warehouse is a repository of an organization's electronically stored data, they are designed to facilitate reporting and analysis (Inmon, 1995). However, the means to retrieve and analyze data, to extract, transform and load data, and to manage the data dictionary are also considered essential components of a data warehousing system. Many references to data warehousing use this broader context. Thus, an expanded definition for data warehousing includes business intelligence tools. In essence, the data warehousing concept was intended to provide an architectural model for the flow of data from operational systems to decision making environments (Wu et al., 2001). Data warehouses are computer based information systems that are home for "secondhand" data that originated from either another application or from an external system or source. Data warehouses are read-only, integrated databases designed to answer comparative and "what if" questions (Giorgini et al., 2008). Unlike operational databases that are set up to handle transactions and that are kept up to date as of the last transaction, data warehouses are analytical, subject-oriented and are structured to aggregate transactions (Al-Debei, 2011).

Moreover, unlike Online Transaction Processing (OLTP), DW is subject-oriented, integrated, non-volatile, and time variant, non-updatable collection of data (See table 2) to support management decision-making processes and business intelligence (Inmon, 2002). DWs contain cleaned, aggregated, consolidated large volumes of data that is accumulated to support multidimensional analysis. Inmon (1995) states that the data warehouse is: "Subject-oriented: The data in the data warehouse is organized so that all the data elements relating to the same real-world event or object are linked together; Time-variant: The changes to the data are tracked and recorded so that reports can be produced showing changes over time; Non-volatile: Data is never over-written or deleted - once committed, the data is static, read-only, and retained for future reporting; Integrated: The data warehouse contains data from most or all of an organization's operational systems and this data is made consistent".

Table 1. Characteristics of Data in a DW (Ang and Teo, 2000)

| Characteristics of Data | Brief description |
|---|---|
| Subject-oriented | Data are grouped by subjects. For example, data on customers are grouped and stored as an interrelated set. |
| Integrated | Data are stored in a globally consistent format. This implies cleansing the data so that data have consistent naming conventions and physical attributes. |
| Time-Variant | Data captured are for long-term use often 5–10 years. So they are captured in a series of snapshots. |
| Non-volatile | Once data at a particular time, say t1, are captured and stored, their attributes are preserved. |

However, after considering the various attributes and concepts of data warehousing systems, a broad definition of a data warehouse can be the following: A data warehouse is a structured extensible environment designed for the analysis of non-volatile data, logically and physically transformed from multiple source applications to align with business applications, updated and maintained for a long time period, and summarized for quick analysis. DW is a data repository which is relevant to the management of an organization and from which the needed information and knowledge to effectively manage the organization are emerged (Watson, 2001).

Table 2. Data Warehouse Definitions

| | Source | Definition |
|---|---|---|
| 1. | Mc Fadden (1996, p. 121). | A collection of integrated, subject oriented databases designed to support the DSS function where each unit of data is relevant to some moment in time. |
| 2. | Subramanian et al., (1997, p.100). | Subject Oriented, integrated, time variant, nonvolatile sets of data in support of management decision making process. |
| 3. | Srivastava and Chen (1999, p. 118). | Subject Oriented, integrated, time variant, nonvolatile sets of data in support of management decision making process. |
| 4. | Samtani et al., (1999, 81). | An integrated repository that stores information which may originate from multiple, possibly heterogeneous operational or legacy data sources. |

| 5. | Ang and Teo (2000, 14). | A repository of summarized data (current as well as historical) assembled in a simplified format tailored for easy end user access. |
| --- | --- | --- |
| 6. | Chen et al., (2000, p. 103). | Subject Oriented, integrated, time variant, nonvolatile collection of data organized to support management needs.# |
| 7. | Watson (2001) | A repository into which are placed all data relevant to the management of an organization and from which emerge the information and knowledge needed to effectively manage the organization. |
| 8. | Bruckner et al., (2001, p. 329). | A common queryable source of data for analysis purposes, which is primarily used as support for decision processes. |
| 9. | Chau et al., (2002, p. 214). | A read only analytical database that is used as the foundation of a decision support system"; "A global repository that stores pre-processed queries on data, which reside in multiple, possibly heterogenous, operational query base for making effective decisions. |
| 10. | Inmon (2002). | Subject Oriented, integrated, time variant, non volatile collection of data in support of management's decisions. |
| 11. | Watson et al., (2004, p. 435). | A repository of data that can be used to support queries, reporting, online analytical processing (OLAP), DSS/EIS, and data mining. |
| 12. | Tseng and Chou (2006, p. 727). | A multi dimensional analyses of cumulated historical business data to help contemporary administrative decision making. |
| 13. | Park (2006, p. 52). | An IT infrastructure that provides appropriate data and tools to support decision makers with a unique opportunity to improve the IT infrastructure. |
| 14. | March and Hevner (2007, p. 1031). | A repository of intelligence from which business intelligence can be derived. |

### 2.1.1 Reasons Underlying the Implementation of Data Warehouses

A wide variety of tangible and intangible benefits (See Figure 1) can be gained from DWs' applications (Watson et al., 2002; Al-Debei, 2011). Initially, Data warehousing is viewed as a way by which business organization could solve the problems associated to their independently legacy systems which often contains inaccurate, duplicate, and dissimilar data about the same entity (Grant, 2003). DW technology can help managers make more effective decisions (Griffin, 1998) by providing them with suitable information which is fundamentally different from the type of information that businesses use in their day-to-day operations (Summer and Ali, 1996).

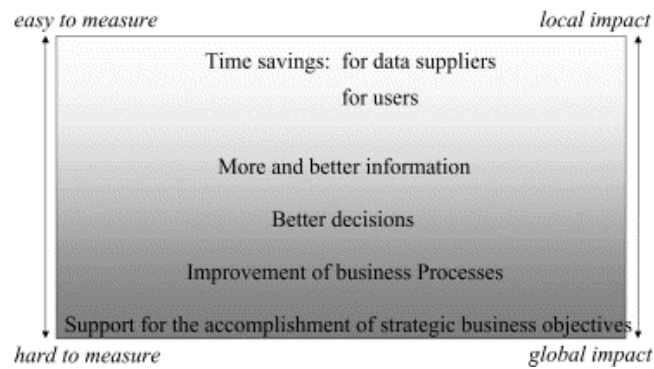

Figure 1. DW Benefits (Adopted from Watson et al., 2002)

DWs have meant to support managers with answers to important business questions that require analytics such as pivoting, drill-downs, roll-ups, aggregations and data slicing and dicing (Ramamurthy et al., 2008). DW allows a business organization to manipulate a great deal of data in ways that are useful to it, such as: cleansing, organizing, describing, summarizing and storing large volumes of data to be transformed, analyzed and reported (Griffin, 1998). Moreover, all levels of management decision-making processes are supported by DW. Alshawi et al. (2003) have elaborated how a DW could provide useful and valuable information and knowledge at a strategic, management control, knowledge and operational levels. DW offers effective service data management and data delivery processes by expanding stovepipe knowledge into cross-functional integrative business intelligence (Shin, 2003). Accordingly, Boar (1997) argues that organizations could compete better by having the ability to learn from the past, to analyze current situations, and to predict the future scenarios. Arnott and Pervan (2008) argue that data warehousing provides the large scale IT infrastructure for contemporary decision support and business intelligence. They argue that the main reasons behind that is the use of Multi-Dimensional Data Model "MDDM" or cubes (Kimball, 1996), which organizes large data sets in ways that are meaningful to managers besides being relatively easy to query and analyze. It has been proved that MDDM is the most suitable for many analytical processes such as data mining, OLAP, and dashboards that used to analyze data from different angles and distilling it into actionable information run over DWs (Gunnarsson et al., 2007).

Moreover, DW provides a foundation for IS/IT application development (i.e. ERP, CRM) which could provide organizations with strategic competitive advantages (Duncan, 1995; Al-Debei, 2011). Another potential benefit of DW is that using a single data source (DW) reduces data inconsistency and redundancy and may facilitate business process re-engineering at business organizations (Watson and Haley, 1998). Despite the argument that the desire to improve decision-making and business performance has been the fundamental business driver behind data warehousing (Gray and Watson, 1998), experts suggest that handling the pressure to comply with governmental regulations such as Sarbanes-Oxley Act and others external pressures which require a real-time disclosure about business operations is the main reason behind the current growth rate of such initiatives (Frolick and Ariyachandra, 2006). Consequently, some of the benefits that a data warehouse provides are as follows (Yang, 1998; Caldeira, 2008):

- A data warehouse provides a common data model for all data of interest regardless of the data's source. This makes it easier to report and analyze information than it would be if multiple data models were used to retrieve information such as sales invoices, order receipts, general ledger charges, etc.

-  Prior to loading data into the data warehouse, inconsistencies are identified and resolved. This greatly simplifies reporting and analysis.

- Information in the data warehouse is under the control of data warehouse users so that, even if the source system data is purged over time, the information in the warehouse can be stored safely for extended periods of time.

- Because they are separate from operational systems, data warehouses provide retrieval of data without slowing down operational systems.

- Data warehouses can work in conjunction with and, hence, enhance the value of operational business applications, notably customer relationship management (CRM) systems.

- Data warehouses facilitate decision support system applications such as trend reports (e.g., the items with the most sales in a particular area within the last two years), exception reports, and reports that show actual performance versus goals.

### 2.2 Information Systems (IS) Success Model

The success of information systems is behind the success of any organization. And because success can be assessed at various levels and from multidimensional perspectives, that explains the continuous emergence of the IS success models. Early attempts of portraying metrics contributing to information systems success have been conceptually frame worked by DeLone and Mclean, 1992. This literature-based initiation has been derived to redefine the fuzziness of information systems success by then, and for the need of a more comprehensive measurement instrument (Seddon and Kiew, 1997; Heo and Han, 2003; DeLone and McLean, 2002, 2004; Roldan and Leal, 2003; Wu and Wang, 2006; Petter et al., 2008; Al-Debei et al., 2013). Therefore, the critical understanding of the nature of IS success grasped attention for more consistent success metrics as researchers began to propose further reformulations and modifications for this widely accepted model within different contexts (i.e DeLone and McLean, 2002, 2003; Roldan and Leal, 2003; Wu and Wang, 2006; Petter et al., 2008; Jalal and Al-Debei, 2010; Al-Debei, 2013). Thus, synthesis of previous research on the IS success model resulted in a deeper understanding of the elusive definition of IS success that was the main purpose behind DeLone and McLean breakthrough (DeLone and McLean, 2002, 2003; Wu and Wang, 2006). Therefore, the motivation behind this study as well as other studies investigating the measures of IS success differently relates to what DeLone and McLean commented: "this success model clearly needs further development and validation before it could serve as a basis for the selection of appropriate IS measures". And they themselves proposing an updated model (which is shown below) assure that.

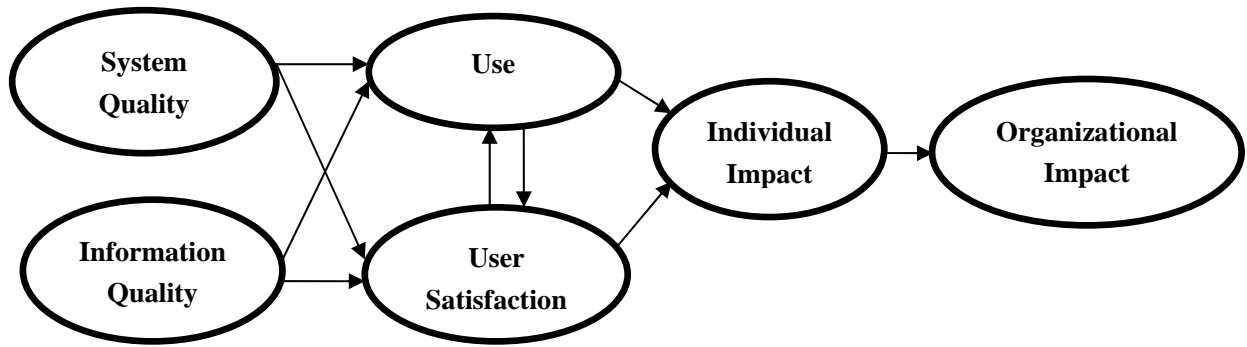

Figure 2. DeLone and McLean IS Success Model (1992)

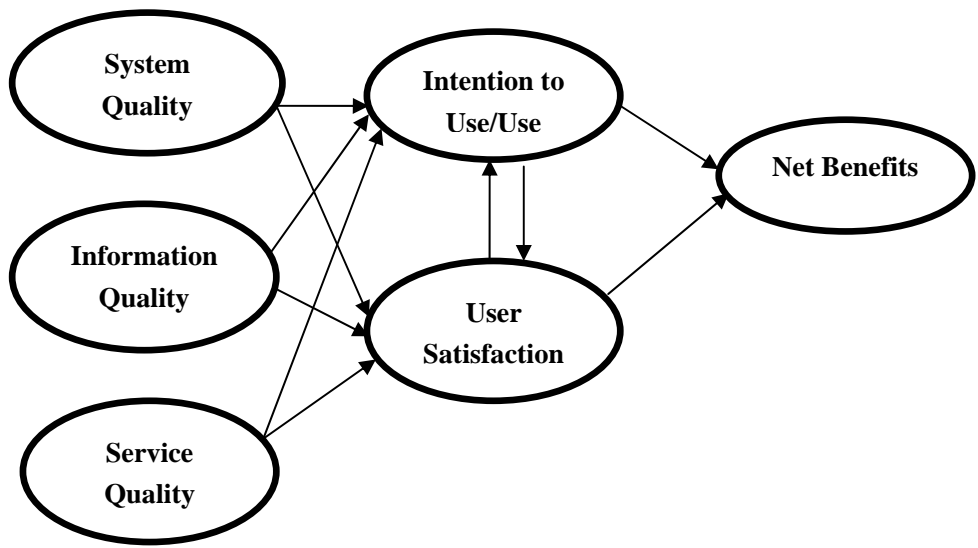

Figure 3. DeLone and McLean's Updated IS Success Model (2002)

The D&M model consists of measures of: System Quality, Information Quality, Service Quality, Intention to Use/Use whether it is a voluntary use or a compulsory one, User Satisfaction, and Net Benefits which is referred to "net" due to the positive or negative possibility of outcomes. On the other hand, despite its wide acceptance, debates on the DeLone and McLean model addressed the relationships of the right hand side of the framework. For example, researchers critically reviewed on whether System Use is to be considered a suitable IS success measure, nevertheless, DeLone and McLean explained the need to study this measure within a deeper sense, considering the extent, nature, appropriateness and quality of it (Wu and Wang, 2006). Additionally, Net Benefits is most widely perceived in terms of measuring monetary values and costs as a result of using the information system, however the modified version of the D&M suggests that this measure should be objectively tackled to measure the intangible perceived usefulness of the system (Wixom and Watson, 2001). Moreover, the linear causality of relationships between System Use, User Satisfaction, Individual Impact and Organizational Impact has not been substantially confirmed. To give a brief example, Seddon (1997) was one of the first to contend that the D&M model introduced confusion because it mixed causal and process explanation of IS success. Seddon argues that System Use does not cause impacts and benefits, it reflects a behavior that would be a consequence of IS success, and this as well applies to User Satisfaction measure. Other argued as well that the relationship between User Satisfaction and System Use is not causal, they suggest that User Satisfaction causes System Use and not vice versa (Wu and Wang, 2006). Furthermore, this study recommends the D&M model as a base for further empirical and theoretical research through a combination of the technological and human elements as considerable measures to success and tends to evaluate that more thoroughly.

## 3. Proposing a Data Warehouse System Success (DWSS) Framework

Indeed, the weight of this research comes from the increased importance of the data warehouses due to the enormous volumes of unmanaged data faced by organizations nowadays. Furthermore, areas studied by this research have been narrowly examined before, as DW systems receive a great deal of attention, yet many questions remain unexplored. Scarce literature concentrated on evaluating DW systems in the developing countries in terms of measuring and evaluating their success. Even though researchers did put lots of effort into finding the most comprehensive IS success framework, there is still a call for building the most compatible one, that not just supports what exists in literature, but also reunites the extant models of IS awareness, adoption, acceptance and success. The exemplified model of this research aims to over hall the gaps of the missing technical, managerial and functional aspects of the traditional information systems success models in a way that revitalizes the DW success model as a whole.

Despite that the context addressed in this research has been widely investigated by practitioners and academics, the originality of this work comes from the conjunction of successful IS theories; theories of Information Systems Implementation (ISI) models, and Delone and McLean's (2002) IS success model. The framework signifies not only the success of data warehouses as systems, but also explains the success of that in terms of achieving strategic business intelligence for decision making purposes in a developing country. From a holistic point of view, the researcher differentiates this work by building up a comprehensive Organizational/Functional model.

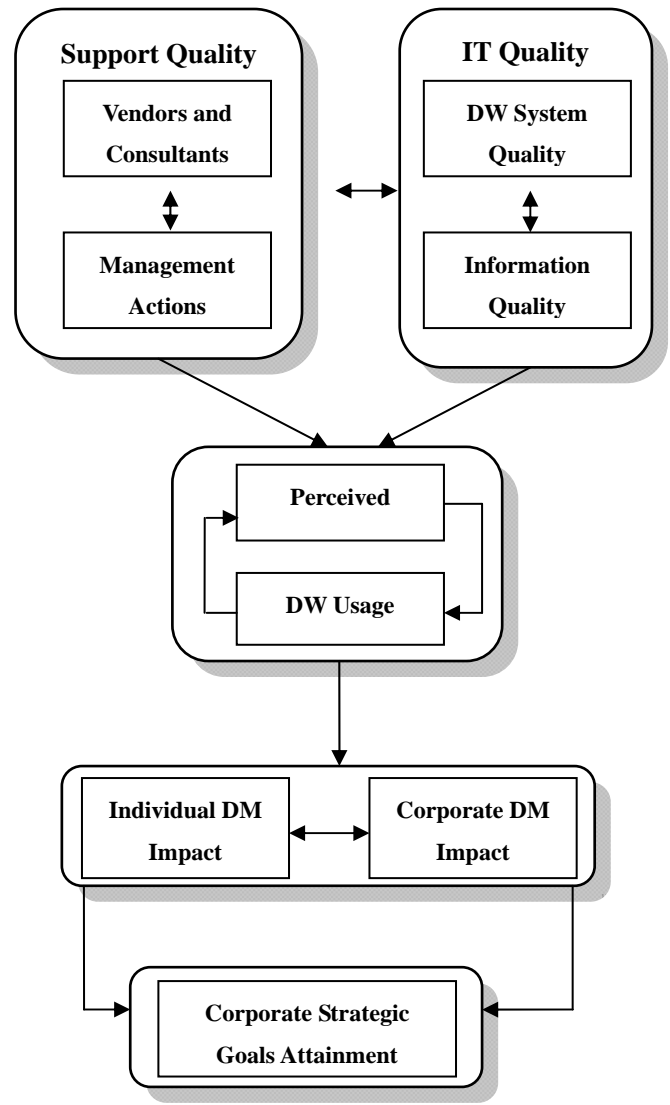

Figure 4. DW Success Framework

The D&M success model has proposed causal relationships of metrics measuring success; however, it lacks the understanding of the importance of some supportive measures related to the technology itself, as well as the users of this system, such as actions held by the organization. As information systems are described as the interaction between people, processes and organizations, there is a need to combine what so called pre-implementation factors with the post implementation ones. The framework of this research illustrates that through the following interrelated levels: Level 1: "DW Success Inputs" is divided into two categories; Support Quality: services and support offered by external entities (vendors and consultants) as well as the internal ones (management); Technology Quality: measures related to the technology itself such as the characteristics of the system and the information it captures (data, software, hardware). Level 2 demonstrates the "DW Use": issues of the usage of the system and the perceived utility out of it. Level 3 exemplifies the factors augmenting "BI" such as individual DM impact that leads to corporate DM in turn; these two measures are enhanced throughout the satisfaction of the DW users. At last but not least, the "DW success outputs" in Level 4 characterized by the success of all the above levels in reaching out to the overall strategic goals and attaining them as effectively and efficiently as possible.

## 4. Discussion

Creating a DW has itself proved to be difficult and problematic, and it is highly perceived as high-risk/high-return initiatives (Watson et al., 2002). The researcher analysis reveals that despite agreement in the information system literature on the importance of DW to an organization success through enhancing its decision-making quality, the attainment of such business intelligence based on it is still poorly proved empirically. Moreover and although the wide variety of motivators mentioned in the IS literature for developing DWs, complying with the governmental regulations is the most conceivable reason for which organizations are building their DWs. DW is highly recognized as an infrastructure; many applications can run over it such as CRM and DSS systems. On the other hand, many techniques, such as data mining, OLAP and dashboards have been rising to prominence to extract business intelligence from DWs. Furthermore, DWs meant to be used by managers since they support decision-making process. Nevertheless, these techniques are still not very effective and are highly perceived as technically oriented by the end-users. However, DW has experienced relatively high failure rates and its spread and/or use has been to some extent limited. Perhaps due to the facts that designing and developing a DW is a risky, costly and complex process. It requires a huge amount of money as an investment, spans over years, and needs a wide variety of technical and managerial skills. Generally speaking, technology is shaped by its social contexts. Hence, their consistent interaction is the key determinant of DW success. Nevertheless and despite the technical complexity of DW design and implementation, social/cultural and organizational factors are the most cited reasons behind DW failures. From these preliminary insights and conclusions, it is recommended to focus on areas such as improving business intelligence techniques in terms of user-friendliness and effectiveness as well as the integration of semi-structured and unstructured data from the knowledge management perspective.

This study has sought to highlight the managerial and technical need for tracking the success of IS investments. The black box nature of IS success has been the reason behind the continuous interest in this field. The framework proposed in this research is a conjunction of an extension of the D&M view of IS success along with other critical implementation considerations (i.e. the support quality of management actions, training sessions, organizational culture, consultancy, etc). Noticeably, there is still the need for a careful and balanced discussion for capturing strategic business value of BI through leveraging DW investments. As a result, the significance novelty of this piece of work seeks to exploit what is beyond the traditional success of DW systems. First, contrary to much of the extant literature in IS success, the DWSS model combined the measures of any sustainable information system, and not any system would endure without the continuous actions and involvement of the organization implementing it itself. That what has been missing within research that adopted and extended the D&M IS model. Logically, the IS literature has not made such a contribution to this field, adopting the same model within various contexts is not much of a value. In conclusion, the folds behind this research are: The visible landmarks of the D&M model, yet the need for further justifications and enhancements as DeLone and McLean (1992) articulated, has encouraged the researcher to represent a more comprehensive framework that combines the functional aspects (i.e. Seddon and Kiew, 1996; Gable et al., 2003; Heo and Han, 2003; Wu and Wang, 2006; Petter et al., 2008) along with the supportive organizational aspects (i.e. McFadden, 1996; Ang and Teo, 2000; Watson et al., 2002; Hwang et al., 2004; Ramamurthy et al., 2008) of a successful information system. This framework is to be adopted by organizations seeking the best out of managing their informational assets, and yet being capable of strategically obtain business intelligence. Additionally, there has been a lack of successful models portraying the DW projects for the purpose of maintaining strategic decision making in developing countries. To be more accurate, literature of DW projects shed the lights on factors of successful adoption and implementation separately from those relating to maintaining it on the long run (i.e. Bruckner et al., 2001; Wixom and Watson, 2001; Hwang et al., 2004; Solomon, 2005; Ramamurthy et al., 2008).

## 5. Conclusions

This study offers future implications for theory and practice. From the theory perspective, the research framework has attributed that the measures of success originated by DeLone and McLean (1992, 2002, 2003, and 2004) are not sufficient for revitalizing the success of data warehouses. The framework test what previous research was deficient in and that is considering IS implementation factors as critical as the traditional IS success measures. From the practical perspective, this research shows the importance of not neglecting the IS implementation factors in assessing and measuring its success, it embodies and proves that no single IS can be successful and sustain this success without the continuous involvement of other parties rather than users of the system. Thus, DW project teams should deliberately and sensibly address the ongoing DW system, and ensure high quality characteristics, and this can't be achievable without the presence of top management, consultancy, vendors, etc. In sum, this research underscores these points and highlights the importance of further model enhancements and justifications in the developing countries which still remains high on the researcher's agenda, as the model is tested through different stages on different time scales.

To remain high on the future agenda is that the proposed framework could be modified to serve as a contingency model in future research, as no single blueprint for success is guaranteed. A future framework that would recognize situation specific factors for each information system and build the ideal model upon that, taking into consideration external variables along with the external ones (i.e. culture, policies, regulations).

## References

Al-Debei, M.M. (2011). Data Warehouse as a Backbone for Business Intelligence. *European Journal of Economics, Finance, and Administrative Sciences*, *33*, 153-166.

Al-Debei, M.M., Jalal, D., & Al-Lozi, E. (2013). Measuring web portals success: a respecification and validation of the DeLone and McLean information systems success model'. *International Journal of Business Information Systems*, Forthcoming.

Al-Debei, M.M. (2013). The quality and acceptance of websites: an empirical investigation in the context of higher education. *International Journal of Business Information Systems*, Forthcoming.

Alshawi, S., Saez-Pujol, I., & Irani, Z. (2003). Data warehousing in decision support for pharmaceutical R & D supply chain. *International Journal of Information Management*, 23, 259-268. http://dx.doi.org/10.1016/S0268-4012(03)00028-8

Ang, J., & Teo, T. (2000). Management Issues in Data Warehousing: Insights from the Housing and Development Board. *Decision Support Systems, 29,* 11-20. http://dx.doi.org/10.1016/S0167-9236(99)00085-8

Arnott, D., & Pervan, G. (2008). Eight Key Issues for the Decision Support Systems Discipline. *Decision Support Systems, 44*, 657-672. http://dx.doi.org/10.1016/j.dss.2007.09.003

Boar, B. (1996). Understanding data warehousing strategically. In R. Barquin, & H. Edelstein (Ed.). *Building, using and managing the data warehouse*, 277–299, New Jersey: Prentice-Hall PTR.

Bruckner, R., List, B., & Schiefer, J. (2001). Developing Requirements for Data Warehouse Systems with Use Cases. *Seventh America's Conference on Information Systems,* 329-335.

Caldeira, C. (2008). *Data Warehousing - Conceitos e Modelos*. Edições Sílabo. ISBN 978-972-618-479-9.

Chau, K., Cao, Y., Anson, M., & Zhang, J. (2002). Application of Data Warehouse and Decision Support System in Construction Management. *Automation in Construction, 12*, 213-224. http://dx.doi.org/10.1016/S0926-5805(02)00087-0

Chen, L., Soliman, K., Mao, E., & Frolick, M. (2000). Measuring User Satisfaction with Data Warehouses: An Exploratory Study. *Information and Management, 37*, 103-110. http://dx.doi.org/10.1016/S0378-7206(99)00042-7

DeLone, W., & McLean, E. (1992). Information Systems Success: The Quest for the Dependent Variable. *Information Systems Research,* 3, 60-95. http://dx.doi.org/10.1287/isre.3.1.60

DeLone, W., & McLean, E. (2002). Information Systems Success Revisited. *Proceedings of the 35th Hawaii International Conference on System Sciences,* 1-11. http://dx.doi.org/10.1109/HICSS.2002.994345

DeLone, W., & McLean, E. (2003). The DeLone and McLean Model of Information Systems Success: A Ten Year Update. *Journal of Management Information Systems, 19*, 9-30.

DeLone, W., & McLean, E. (2004). Measuring e-Commerce Success: Applying the DeLone and McLean Information Systems Success Model. *International Journal of Electronic*

*Commerce, 9*, 31-47.

Frolick, M., & Ariyachandra, T. (2006). Critical Success Factors in Business Performance Management-Striving for Success. *Information Systems Management, 25,* 113-120.

Gable, G., Sedera, D., & Chan, T. (2003). Enterprise Systems Success: A Measurement Model. *Twenty Forth International Conference on Information Systems,* 576-591.

Giorgini, P., Rizzi, S., & Garzetti, M. (2008). GRAnD: A Goal-Oriented approach to requirement analysis in Data Warehouses. *Decision Support Systems, 45,* 4-21. http://dx.doi.org/10.1016/j.dss.2006.12.001

Gray, P., & Waston, H. J. (1998). Present and Future Directions in Data Warehousing. *Database for Advances in Information Systems*, *29,* 83-90. http://dx.doi.org/10.1145/313310.313345

Griffin, R. K. (1998). Data Warehousing. *Cornell Hospitality Quarterly, 39,* 28-35.

Gunnarsson, C. L., Walker, M. M., Walatka, V., & Swann, K. (2007). Papers Lessons Learned: A case study using data mining in the newspaper industry. *Database Marketing & Customer Strategy Management*, *14,* 271-280. http://dx.doi.org/10.1057/palgrave.dbm.3250058

Heo, J., & Han, I. (2003). Performance Measure of Information Systems (IS) in Evolving Computing Environments: An Empirical Investigation. *Information and Management, 40,* 243-256. http://dx.doi.org/10.1016/S0378-7206(02)00007-1

Hwang, H., Ku, C., Yen, D., & Cheng, C. (2004). Critical Factors Influencing the Adoption of Data Warehouse Technology: A study of the Banking Industry in Taiwan. *Decision Support Systems, 37*, 1-21. http://dx.doi.org/10.1016/S0167-9236(02)00191-4

Inmon, W.H. (1995). *Tech Topic: What is a Data Warehouse?*. Prism Solutions, *1*, 3-11.

Inmon, W.H. (2002). *Building the Data Warehouse*. Wiley Computer Publishing, 3$^{rd}$ edition, 1-343.

Jalal, D., & Al-Debei, M.M. (2012). Portals and Task Innovation: A Theoretical Framework Founded on Business Intelligence Thinking. *Proceedings of the Eleventh Annual International Conference on Business Intelligence and Knowledge Economy*, Al-Zaytoonah University, Amman, Jordan.

Jennex, M., & Olfman, L. (2003). A Knowledge Management Success Model: An Extension of DeLone and McLean's IS Success Model. *Association for Information Systems, 11*, 2529-2539.

Kimball, R. (1996). *The data warehouse toolkit.* Practical techniques for building dimensional data warehouses, Wiley, New York.

March, S., & Hevner, A. (2007). integrated Decision Support Systems: A Data Warehousing Perspective. *Decision Support Systems, 43*, 1031-1043. http://dx.doi.org/10.1016/j.dss.2005.05.029

McFadden, F. (1996). Data Warehouse for EIS: Some Issues and Impacts. *Proceedings of the 29th Annual Hawaii International Conference on System Sciences,* 120-335.

Park, Y. T. (2006). An Empirical Investigation of the Effects of Data Warehousing on Decision Performance. *Information and Management, 43*, 51-61. http://dx.doi.org/10.1016/j.im.2005.03.001

Petter, S., DeLone, W., & McLean, E. (2008). Measuring Information Systems Success: Models, Dimensions, Measures, and Interrelationships. *European Journal of Information Systems, 17,* 236-263. http://dx.doi.org/10.1057/ejis.2008.15

Ramamurthy, K.R., Sen, A., & Sinha, A.P. (2008). An Empirical Investigation of the Key Determinants of Data Warehouse Adoption. *Decision Support Systems, 44*, 817-841. http://dx.doi.org/10.1016/j.dss.2007.10.006

Roldan, J., & Leal, A. (2003). A Validation Test of an Adaptation of the DeLone and McLean's Model in the Spanish EIS Field. *Critical Reflections on Information Systems: A Systematic Approach,* 66-84.

Saarinen, T. (1996). An Expanded Instrument for Evaluating Information System Success. *Information and Management, 31,* 103-118. http://dx.doi.org/10.1016/S0378-7206(96)01075-0

Samtani, S., Mohania, M., Kumar, V., & Kambayashi, Y. (1999). Recent Advances and Research Problems in Data Warehousing, *Springer Verlag Berlin Heidelberg,* 81-92.

Sean, M., Rouse, A., & Beaumont, N. (2007). Explaining and Predicting Information Systems Acceptance and Success: An Integrative Model. *ECIS*, 1356-1367.

Seddon, P., & Kiew, M. (1997). A Partial Test and Development of Delone and McLean's Model of IS Success. *AJIS, 4*, 90-109.

Seddon, P. (1997). A Respecification and Extention of the DeLone and McLean Model of IS Success. *Information Systems Research, 8,* 240-253. http://dx.doi.org/10.1287/isre.8.3.240

Shin, B. (2003). An Exploratory Investigation of System Success Factors in Data Warehousing. *Journal of the Association for Information Systems, 4*, 141-170.

Solomon, M. (2005). Ensuring a Successful Data Warehouse Initiative. *Enterprise Systems Management Journal, 22,* 26-36.

Stake, R.E. (1995). *The art of case study research'.* Thousand Oaks, CA: Sage.

Strivastava, J., & Chen, P. (1999). Warehouse Creation-A Potential Roadblock to Data Warehousing. *IEEE Transactions on Knowledge and Data Engineering, 11*, 118-126. http://dx.doi.org/10.1109/69.755620

Subramanian, A., Smith, L., Nelson, A., Campbell, J., & Bird, D. (1997). Strategic Planning for Data Warehousing. *Information and Management, 31,* 99-113. http://dx.doi.org/10.1016/S0378-7206(97)00040-2

Summer, E., & Ali, D. (1996). A Practical Guide for Implementing Data Warehousing. *Computers Ind. Engng, 31,* 307-310. http://dx.doi.org/10.1016/0360-8352(96)00137-4

Tellis, W. (1997). Introduction to Case Study. *The Qualitative Report, 3,* 33-45.

Tseng, F., & Chou, A., (2006). The Concept of Document Warehousing for multi-dimensional modeling of Textual-Based Business Intelligence. *Decision Support Systems, 42,* 727-744. http://dx.doi.org/10.1016/j.dss.2005.02.011

Watson, H., Goodhue, D., & Wixom, B. (2002). The Benefits of Data Warehousing: Why some Organizations Realize Exceptional Payoffs. *Information and Management, 39,* 491-502. http://dx.doi.org/10.1016/S0378-7206(01)00120-3

Watson, H. (2001). Recent Developments in Data Warehousing. *Association for Information Systems "AMCIS 2001 Proceedings",* 2289-2292.

Watson, H., Fuller, C., & Ariyachandra, T. (2004). Data Warehouse Governance: Best Practices at Blue Cross and Blue Shield of North Carolina. *Decision Support Systems, 38,* 435-450. http://dx.doi.org/10.1016/j.dss.2003.06.001

Watson, H., & Haley, B. (1998). A Structural Model of Data Warehousing Success. *Journal of Data Warehousing, 2*, 10-17.

Wixom, B., & Watson, H. (2001). An Empirical Investigation of the Factors Affecting Data Warehousing Success. *MIS Quarterly, 25*, 17-41. http://dx.doi.org/10.2307/3250957

Wu, J., & Wang, Y. (2006). Measuring KMS Success: A Respecification of the DeLone and McLean's Model. *Information and Management, 43*, 728-739. http://dx.doi.org/10.1016/j.im.2006.05.002

Wu, L., Miller, L., & Nilakanta, S. (2001). Design on Data Warehouses using Metadata. *Information and Software Technology, 43*, 109-119. http://dx.doi.org/10.1016/S0950-5849(00)00143-9

Yang, J. (1998) *Ware House Information Prototype at Stanford (WHIPS).* Stanford University.

Yin, R.K. (1984). *Case Study Research: Design and methods.* Newbury Park, CA: Sage.

Yin, R. (1994). *Case Study Research: Design and Methods,* 2nd edition. Beverly Hills, CA: Sage Publishing.

**Copyright Disclaimer**